\documentstyle [11pt]{article}
\begin{document}
\input epsf
\newtheorem{D}{Definition}[section]
\newtheorem{T}{Theorem}[section]

\title{Does Quantum Mechanics Save Free Will?\footnote{ Presented
at the conference "Einstein meets Magritte" , Brussels, 29 May -- 3
June 1995.}}
\author{ L\'aszl\'o E. Szab\'o\thanks{ E-mail: leszabo@ludens.elte.hu}
\\ {\it Institute for Theoretical Physics } \\ {\it E\"otv\"os University,
Budapest } }
\date{ }
\maketitle
\begin{abstract}
According to the widely accepted opinion, classical (statistical) physics
does not support objective indeterminism, since the statistical laws of
classical physics allow a deterministic hidden background, while --- as
Arthur Fine writes polemizing with Gr\"unbaum --- "{\sl the
antilibertarian position finds little room to breathe in a statistical world
if we take laws of the quantum theory as exemplars of the statistical
laws in such a world. So, it appears that, contrary to what Gr\"unbaum
claims, the libertarians' 'could have done otherwise' does indeed find
support from indeterminism if we take the indeterministic laws to be of
the sort found in the quantum theory.}"

In this paper I will show that, quite the contrary, quantum mechanics
does not save free will. For instance, the EPR experiments are
compatible with a deterministic world. They admit a deterministic
local hidden parameter description if  the deterministic model is
'allowed' to describe not only the measurement outcomes, but also the
outcomes of the 'decisions' whether this or that measurement will be
performed. So, the derivation of the freedom of the will from quantum
mechanics is a tautology: from the assumption that the world is
indeterministic it is derived that the world cannot be deterministic.
\end{abstract}

\paragraph{1.} To avoid a kind of "Comedy of Errors" I should specify
in a very strict way what I mean under "free will" and how it relates to
"indeterminism". However, the full analysis of these two notions would
be outside the scope of this paper. The only thing I can do is to
describe that particular aspect of the question on which we focus our
attention.
\begin{center}
\framebox[11cm][c]{\vbox{ \centerline{ \epsfxsize=10.5cm \epsfbox{
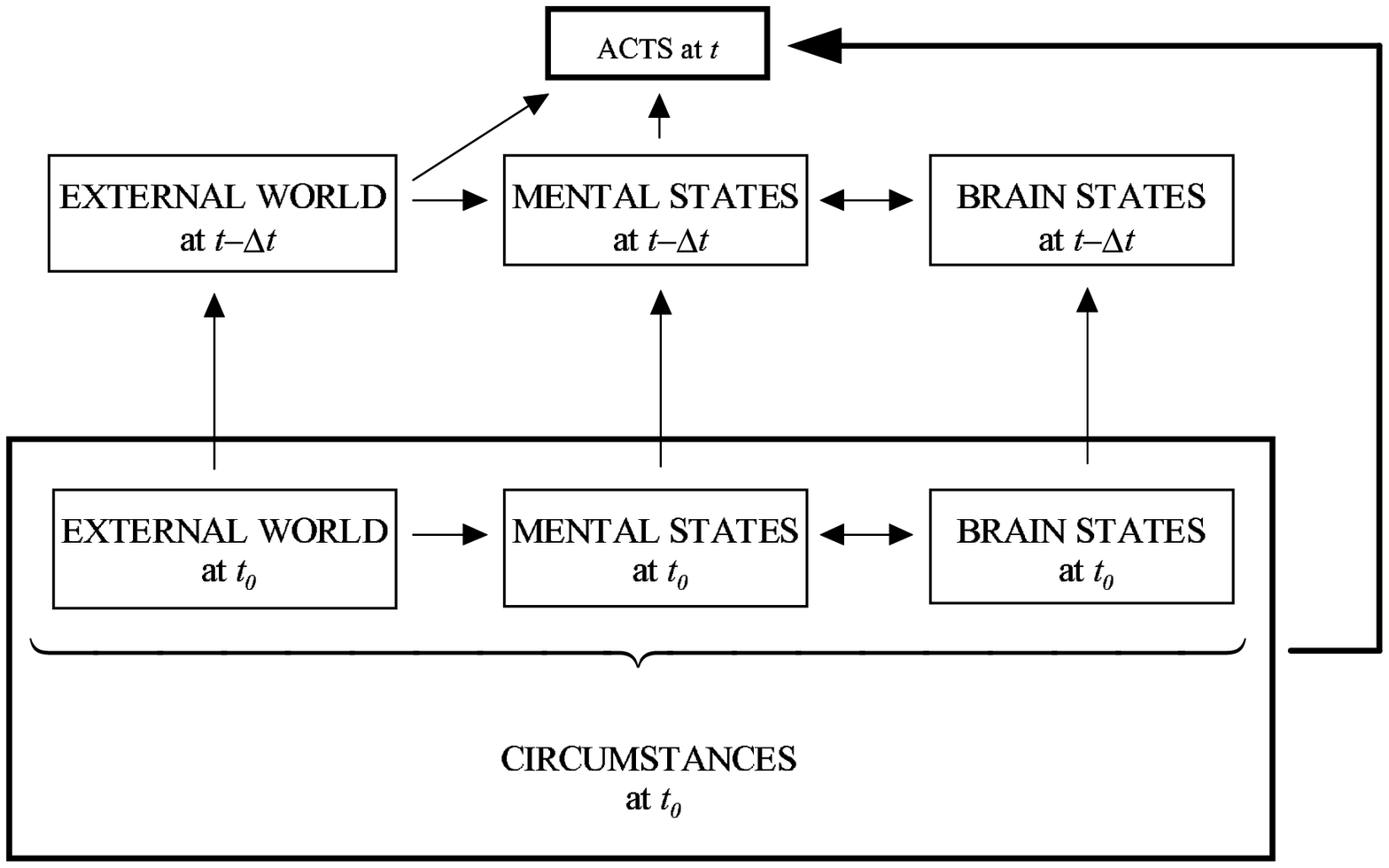}}  \par {\it Figure 1.}}}
\end{center}

\noindent
Each arrow in Fig.~1 symbolizes a particular item within the problem
of the freedom of human action: how the mental states determine
human actions, how the mental state at time $t$  is determined by the
mental state at an earlier time $t_{0}$, are the brain processes
deterministic or indeterministic, how the mental states and the brain
states are related, etc. Without going to these details I would like to
recall three different concepts of the freedom of the will.

\begin{itemize}
\item[A)] According to the commonsense understanding, the freedom
of the will means that someone can do what he thinks to do. This
concept of freedom is connected to the problem of relation between the
acts and mind states.
\item[B)] According to another concept, free will means the subjective
feeling of that someone's future decision or future thought is open for
himself. In this way the freedom of the will relates to the problem of
auto-predictability of an evolving system.
\end{itemize}

\noindent
These first two interpretations of free will {\it are entirely compatible
with determinism}:

\noindent
In case A) a deterministic world even helps to feel a real correlation
between someone's action and thoughts.

\noindent
In case B), as MacKay (1967) proved by applying Popper's (1982)
result concerning the limitations governing auto-predictability of a
deterministically evolving system, the subjective feeling of freedom is
provided even if the world were as mechanical as clockwork.

\noindent
Now I offer to accept, for our purposes, Campbell's definition of the
freedom of the will:

\begin{itemize}
\item[C)] One's action is free only if one could have acted differently
under the same circumstances. In other words, the relation between the
acts at time $t$ and the circumstances (see Fig.~1) at time $t_{0}$  is
{\it not } deterministic.
\end{itemize}

For sake of brevity I shall call this process starting with the given
circumstances at time $t_{0}$   and terminating with the acts at time
$t$  as "choice process".
Free will, as it is understood in C), is not accommodated in a
deterministic universe. In a deterministic world the choice process is
also deterministic, therefore Campbell's freedom of the will, by
definition, {\it does not exist}.

\paragraph{2.} Adolf Gr\"unbaum rejects Campbell's conception of the
freedom of the will as inadequate. He argues that {\it we would not
have this libertarian kind of freedom even if the choice process were
governed by probabilistic laws}. The argument proceeds as follows.

Suppose that an individual $X$  can act at time $t$ in various ways
denoted by $x_{1}, x_{2}, ... x_{n}$, given some fixed circumstances
at time $t_{0}$. Assume we have a probabilistic theory describing his
behavior according to which the probabilities of the different acts are
$0< p\left(x_{1}\right), p\left(x_{2}\right), ... p\left(x_{n}\right) < 1$
(each probability is different from 0 or 1, otherwise the choice process
would be obviously deterministic). One can interpret these probabilities
as relative frequencies of the corresponding acts in two senses, either
by taking a large statistical ensemble of identical $X$-like individuals,
or by repeating many times the "experiment" with the same individual
$X$  under the same circumstances.

{\it Does this probabilistic characterization of the choice process entail
that an individual choice from the possible acts at time} $t$ {\it is not
determined in advance at time} $t_{0}${\it ?}
Gr\"unbaum's answer is, definitely not. Suppose a community, he
argues, which is subject to an indeterministic law according to which,
in a long run, 80 percent of the population will commit a certain kind
of crime. We can hold those members of the community who commit
the crime morally responsible for their behavior only if, as the
libertarian standard would have it, they could have done otherwise.
But, the statistical law does not entitle us to say of an individual who
commits the crime that he could have done otherwise. To be sure, on
the basis of the law we cannot tell which particular individuals in the
community will commit the crime. The law does not specify that. But
this limitation does not entail that, in the very circumstances in which
an individual commits the crime, he could have refrained from so
doing.

However, as Arthur Fine (1993) pointed out, Gr\"unbaum's
argumentation depends on whether, in principle, the probabilistic
model in question admits a deterministic hidden variable theory. And
this is the case only if the probability model is a classical
Kolmogorovian one, "{\sl but as we have come to learn in connection
with foundational studies in the quantum theory, just such
counterfactual distinctions may turn out to have unexpected and
testable consequences}", --- as Fine (1993, p. 553) has rightly
remarked. He turns then to prove "the conflict between
antilibertarianism and the quantum theory" by considering a typical
EPR-type experiment. He concludes: "{\sl If we assume that the
quantum theory is correct in its statistical predictions and we hold to
the reasonable no action-at-a-distance condition involved in the stated
locality principle, then it follows that the statistical laws of the
quantum theory cannot be given an antilibertarian interpretation....
So, it appears that, contrary to what Gr\"unbaum claims, the
libertarians' 'could have done otherwise' does indeed find support from
indeterminism if we take the indeterministic laws to be of the sort
found in the quantum theory.}" (pp. 555-556)

Nonetheless, as Fine himself remarks, "{\sl the conclusion that
Gr\"unbaum draws may turn out to be more robust than the particular
argument he gives for it }".

\paragraph{3.} Let us turn now to a more careful analysis of the
alleged conflict between antilibertarianism and the quantum theory. It
is true that the quantum theory could turn out to be in conflict with
antilibertarianism, but --- as we will see it soon --- {\it it doesn't}.

To decide whether a probability model is classical or not one needs
something more than the probabilities $ p\left(x_{1}\right),
p\left(x_{2}\right), ... p\left(x_{n}\right)$. Until we do not consider the
{\it conjunctions} of events $x_{1}, x_{2}, ... x_{n}$, that is, the
events like "individual $X$  executes acts $x_{i}$  {\it and} $x_{j}$",
the probability model can be regarded as a classical/Kolmogorovian
one admitting deterministic hidden variable theory. So, we must
assume that the probabilistic model provides not only the probabilities
$ p\left(x_{1}\right), p\left(x_{2}\right), ... p\left(x_{n}\right)$ but
also probabilities $p\left( x_{i}\& x_{j}\right) $  for some conjunctions
$ x_{i}\& x_{j}$.

And now we inquire: {\it How should we imagine a probabilistic
description "of the sort found in the quantum theory" for the choice
process?}

Take a simple kind of example. Assume a nurse can offer four different
things to a baby: scrambled eggs, pudding, tomato juice and coke. Each
time she offers one kind of food together with one kind of drink.
\begin{center}
\framebox[9.5cm][c]{\vbox{\centerline{ \epsfxsize=9cm
\epsfbox{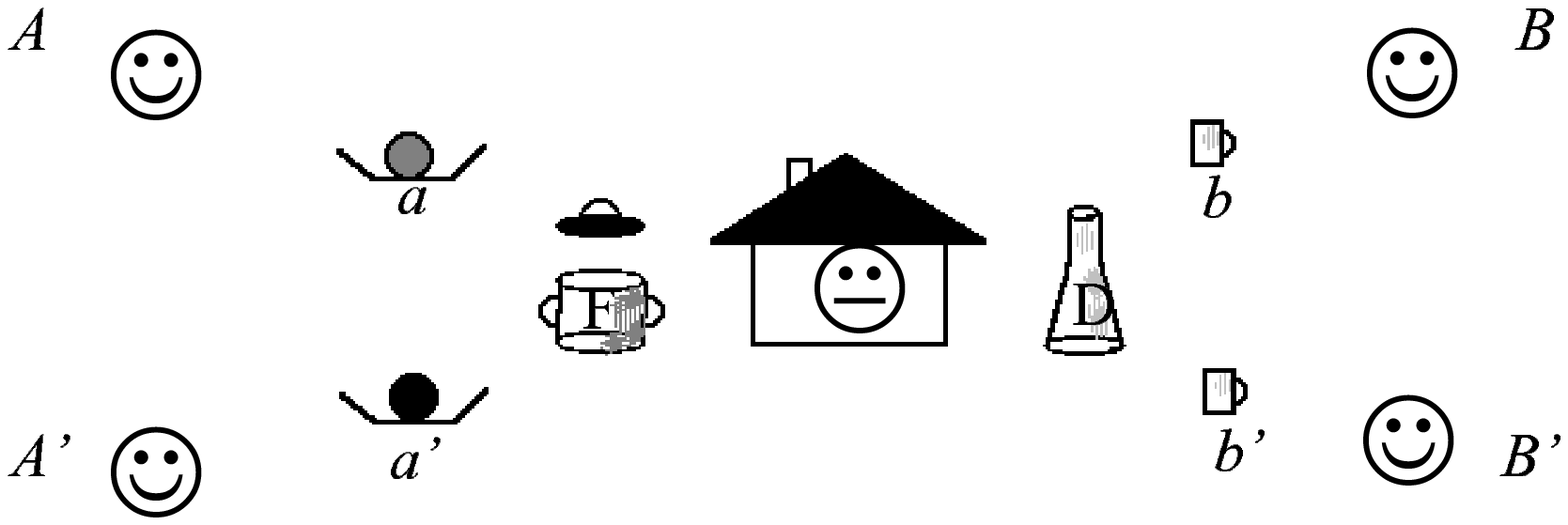}}  \par{\it Figure 2: The quantum baby}}}
\end{center}

The observed events are the followings:
\begin{center}
\begin{tabular}{ rl} 
$A$ :&The baby is eating scrambled eggs\\
$A'$ :&The baby is eating pudding\\
$B$ :&The baby is drinking tomato juice\\
$B'$ :&The baby is drinking coke\\
$a$ :&The nurse offers scrambled eggs\\
$a'$ :&The nurse offers pudding\\
$b$ :&The nurse offers tomato juice\\
$b'$ :&The nurse offers coke
\end{tabular}
\end{center}
The situation is entirely analogous with the Aspect-type
Einstein-Podolsky-Rosen experiment with spin-$\frac{1}{2}$
particles.

The four detectors detect the spin-up events. The two switches are
making choice from sending the particles to the Stern-Gerlach magnets
directed into different directions. The observed events are the
followings:

\begin{center}
\begin{tabular}{rl} 
\leavevmode
$A$ :&The "left particle has spin 'up' into direction ${\bf a} $" detector
beeps\\
$A'$ :&The "left particle has spin 'up' into direction ${\bf a'} $"
detector beeps\\
$B$ :&  The "right particle has spin 'up' into direction ${\bf b} $"
detector beeps\\
$B'$ :& The "right particle has spin 'up' into direction ${\bf b'} $"
detector beeps\\
$a$ :& The left switch selects direction ${\bf a} $ \\
$a'$ :& The left switch selects direction ${\bf a'} $ \\
$b$ :& The right switch selects direction ${\bf b} $ \\
$b'$ :& The right switch selects direction ${\bf b'} $
\end{tabular}
\end{center}
 \begin{center}
\framebox[9.5cm][c]{\vbox{ \centerline{ \epsfxsize=9cm \epsfbox{
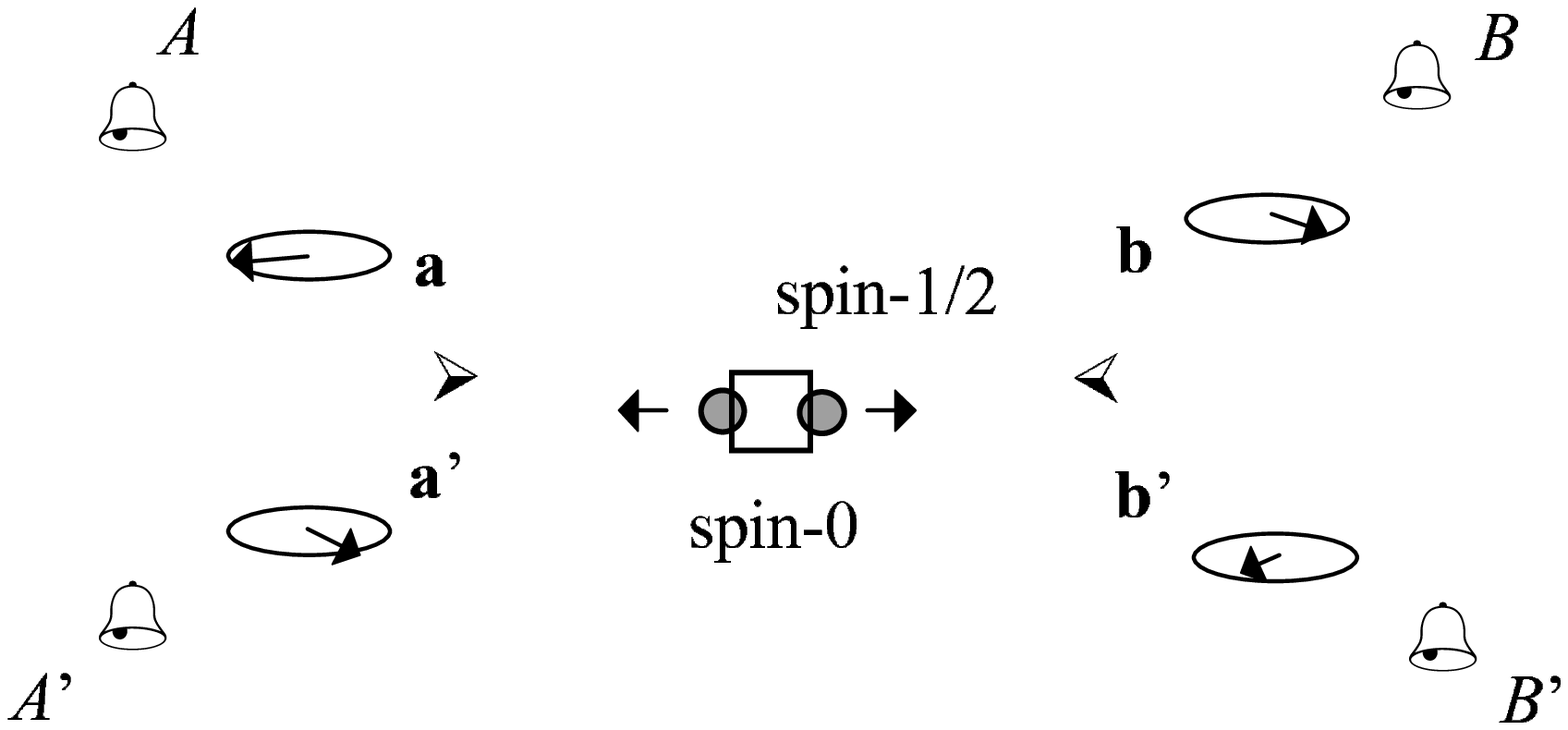}} \par {\it Figure 3: Aspect experiment with
spin-$\frac{1}{2}$  particles }  }}
\end{center}
For the probabilities of these events, in case of $\angle \left( {\bf
a},{\bf a'} \right)= \angle \left( {\bf a'},{\bf b} \right)= \angle \left( {\bf
a},{\bf b'} \right)=120^{\circ } $ and $\angle \left( {\bf b},{\bf a'}
\right)=0$, we have
\begin{eqnarray}
\label{Facts1}
p(A)=p(A')=p(B)=p(B') &=&\frac{1}{4}  \nonumber \\
p(a)=p(a')=p(b)=p(b') &=&\frac{1}{2}  \nonumber \\
p(A\wedge a)=p(A) &=&\frac{1}{4}  \nonumber \\
p(A'\wedge a')=p(A') &=&\frac{1}{4} \\
p(B\wedge b)=p(B) &=&\frac{1}{4}  \nonumber \\
p(B'\wedge b')=p(B') &=&\frac{1}{4}  \nonumber \\
p(A\wedge a')=p(A'\wedge a)=p(B\wedge b')=p(B'\wedge b) &=&0
\nonumber \\
p(A\wedge B)= p(A\wedge B')= p(A'\wedge B')&=&\frac{3}{32}
\nonumber \\
\label{Facts2}
p(A'\wedge B)&=&0  \nonumber \\
p(a\wedge a)= p(b\wedge b')&=&0   \nonumber \\
p(a\wedge b)= p(a\wedge b')= p(a'\wedge b)= p(a'\wedge
b')&=&\frac{1}{4} \\
p(A\wedge b)= p(A\wedge b')= p(A'\wedge b)= p(A'\wedge b')
\nonumber \\ =p(B\wedge a)= p(B\wedge a')= p(B'\wedge a)=
p(B'\wedge a')&=&\frac{1}{8}  \nonumber
\end{eqnarray}
These statistical data agree with quantum mechanical results, in the
sense that
\begin{eqnarray}
\label{qprob}
\frac{p(A\wedge a)}{p(a)}=tr(\hat{W}\hat{A})=\frac{p(A'\wedge
a')}{p(a')}= tr(\hat{W}\hat{A'})\nonumber \\=\frac{p(B\wedge
b)}{p(b)}=
tr(\hat{W}\hat{B})=\frac{p(B'\wedge
b')}{p(b')}=tr(\hat{W}\hat{B'})&=&\frac{1}{2}  \nonumber \\
\frac{p(A\wedge B\wedge a\wedge b)}{p(a\wedge b)}=
\frac{p(A\wedge B)}{p(a\wedge b)}=tr(\hat{W}\hat{A}\hat{B})
\nonumber \\
=\frac{1}{2}\sin^{2}\angle ({\bf a},{\bf b})&=&\frac{3}{8}
\nonumber \\
\frac{p(A\wedge B'\wedge a\wedge b')}{p(a\wedge b')}=
\frac{p(A\wedge B')}{p(a\wedge b')}=tr(\hat{W}\hat{A}\hat{B'})
\nonumber \\
=\frac{1}{2}\sin^{2}\angle ({\bf a},{\bf b'})&=&\frac{3}{8}\\
\frac{p(A'\wedge B\wedge a'\wedge b)}{p(a'\wedge b)}=
\frac{p(A'\wedge B)}{p(a'\wedge b)}=tr(\hat{W}\hat{A'}\hat{B})
\nonumber \\
=\frac{1}{2}\sin^{2}\angle ({\bf a'},{\bf b})&=&0  \nonumber \\
\frac{p(A'\wedge B'\wedge a'\wedge b')}{p(a'\wedge b')}=
\frac{p(A'\wedge B')}{p(a'\wedge b')}=tr(\hat{W}\hat{A'}\hat{B'})
\nonumber \\
=\frac{1}{2}\sin^{2}\angle ({\bf a'},{\bf b'})&=&\frac{3}{8}
\nonumber
\end{eqnarray}
where the outcomes are identified with the following projectors
\begin{eqnarray*}
\hat{A}&=&\hat{P}_{span\left\{ \psi _{+{\bf a}}\otimes \psi _{+{\bf
a}},  \psi _{+{\bf a}}\otimes \psi _{-{\bf a}  }\right\} }\\
\hat{A'}&=&\hat{P}_{span\left\{ \psi _{+{\bf a}}\otimes \psi _{+{\bf
a}},  \psi _{+{\bf a'}}\otimes \psi _{-{\bf a'}  }\right\} }\\
\hat{B}&=&\hat{P}_{span\left\{ \psi _{-{\bf b}}\otimes \psi _{+{\bf
b}},  \psi _{+{\bf b}}\otimes \psi _{+{\bf b}  }\right\} }\\
\hat{B'}&=&\hat{P}_{span\left\{ \psi _{-{\bf b'}}\otimes \psi _{+{\bf
b'}},  \psi _{+{\bf b'}}\otimes \psi _{+{\bf b'}  }\right\} }
\end{eqnarray*}
of the Hilbert space $H^{2}\otimes H^{2}$. The state of the system is
assumed to be represented as $\hat{W}=\hat{P}_{\Psi _{s}}$, where
$\Psi _{s}=\frac{1}{\sqrt{2}}\left( \psi _{+{\bf a}}\otimes \psi _{-{\bf
a} }-\psi _{-{\bf a}}\otimes \psi _{+{\bf a} }\right) $.

Nothing speaks against that the probabilities in (\ref{Facts1}) and
(\ref{Facts2}) describe also the statistics of the baby's behavior if the
nurse makes choice with equal frequencies between the two possible
kinds of food and the two possible kinds of drink.

The numbers in (\ref{qprob}) indeed violate the Clauser-Horne
inequalities. {\it But does it mean that the Aspect experiment as well as
our baby's behavior cannot be accommodated in a deterministic
universe, and consequently quantum mechanics provides the existence
of Campbell's freedom of the will?} Not at all!

As it was shown in Szab\'o 1994 and Szab\'o 1995, quantum mechanics
itself turns out to be reducible, that is, it admits a local deterministic
hidden variable theory. At least, this is proved for the EPR experiments
on which the libertarian quantum-indeterminists so often base their
arguments. A "deterministic universe" includes not only the causal
determination of the measurement outcomes, but it also includes
causally deterministic decision processes of whether this or that
measurement is being performed. No matter whether these decisions
are made by machines like the switches in the Aspect experiment or by
human beings, such processes must be deterministic in a deterministic
world. So, if we ask whether the Aspect experiment admits a
deterministic (and local) hidden variable theory, {\it we must draw into
the consideration the behavior of the switches, too} (Cf. Brans 1988).

It turns out that {\it the whole system together can be deterministic}:
The probabilities in (\ref{Facts1}) and (\ref{Facts2})  form a
correlation vector (see Pitowsky 1989)
\begin{eqnarray}
{\bf p}=\left( p(A)p(A')p(B) \ldots p(b)p(b')p(A\wedge A')p(A\wedge
B)p(A\wedge B') \ldots p(b\wedge b')\right)  \nonumber \\
=\left(
\frac{1}{4}\frac{1}{4}\frac{1}{4}\frac{1}{4}\frac{1}{2}\frac{1}{2}
\frac{1}{2}\frac{1}{2}0\frac{3}{32}\frac{3}{32}\frac{1}{4}\frac{1}
{8}\frac{1}{8}\frac{1}{8}0\frac{3}{32}0\frac{1}{4}\frac{1}{8}
\frac{1}{8}0\frac{1}{8}\frac{1}{8}\frac{1}{4}0\frac{1}{8}\frac{1}{
8}0
\frac{1}{4}0\frac{1}{4}\frac{1}{4}\frac{1}{4}\frac{1}{4}0\right)
\end{eqnarray}
which satisfies Pitowsky's geometric condition: It is in the classical
correlation polytope, ${\bf p}\in C\left( 8, S^{max}_{28} \right)  $,
which is a sufficient condition for reducibility, in the sense that the
probability model in question is Kolmogorovian and admits a local
deterministic hidden parameter theory (see Szab\'o 1995).

The same holds for the baby's choice example. In an entirely
deterministic world not only the baby's behavior is deterministic, but
the nurse's decision is deterministic, too. And {\it the whole  "baby +
nurse" system can be accommodated in a deterministic world}. In other
words, it can be the case that neither the nurse nor the baby has free
will.

You may not like this solution of the problem, arguing that the nurse
must have free will (as well as the switches in the Aspect experiment
must be indeterministic) by assumption. But it should be clear that in
this case the derivation of the freedom of the will from the quantum
theory becomes tautology: if there exists free will then free will exists.
Although, this result has been proved for a particular --- but very
important --- case, one cannot raise any objection against believing that
the same  holds for an arbitrary situation in the quantum theory. And if
it is so, Gr\"unbaum's conclusion really "turns out to be more robust
than the particular argument he gives for it".

\section*{References}
\begin{list}%
{ }{\setlength{\itemindent}{-15pt}
\setlength{\leftmargin}{15pt}}
\item Brans, C. H. 1988: "Bell theorem does not Eliminate fully causal
hidden variables", {\it Int. J. of Theor. Phys.} {\bf 27}  p. 219.
\item Fine, A. 1993:  "Indeterminism and the Freedom of the Will", In
{\it Philosophical Problems of the Internal and External World ---
Essays on the Philosophy of Adolf Gr\"unbaum}, J. Earman, A. I.
Janis, G. J. Massey, N. Rescher, eds., Univ. of Pittsburgh Press /
Universit\"atsverlag Konstanz, p. 551.
\item Gr\"unbaum, A. 1972:  "Free Will and Laws of Human
Behavior", in {\it New Readings in Philosophical Analysis}, H. Feigl,
W. Sellars, K. Lehrer, eds., New York: Appleton-Century-Crofts, p.
614.
\item MacKay, D. M.1967: {\it Freedom of Action in a Mechanistic
Universe}, Cambridge Univ. Press
\item Pitowsky, I. 1989: {\it Quantum Probability --- Quantum Logic},
Lecture Notes in Physics {\bf 321}, Springer, Berlin.
\item Popper, K. R. 1982: {\it The Open Universe --- An Argument for
Indeterminism}, Hutchinson, London.
\item Szab\'o, L. 1994:  "Quantum mechanics in a deterministic
universe", forthcoming in  {\it Quantum Structures '94} (Int. J. of
Theor. Phys.)
\item Szab\'o, L. 1995:  "Is quantum mechanics compatible with an
entirely deterministic universe? Two interpretations of quantum
probabilities", forthcoming in {\it Foundations of Physics Letters.}

\end{list}

\end{document}